# VALIDATION OF OBSERVATIONS OBTAINED WITH A LIQUID MIRROR TELESCOPE BY COMPARISON WITH SLOAN DIGITAL SKY SURVEY OBSERVATIONS

E.F. Borra

Centre d'Optique, Photonique et Lasers

Département de Physique, Université Laval

## ABSTRACT

The results of a search for peculiar astronomical objects using very low resolution spectra obtained with the NASA Orbital Debris Observatory (NODO) 3 meter diameter liquid mirror telescope (LMT) are compared with results of spectra obtained with the Sloan Digital Sky Survey (SDSS). The main purpose of this comparison is to verify whether observations taken with this novel type of telescope are reliable. This comparison is important because LMTs are a novel type of inexpensive telescope that are very useful for astronomical surveys, particularly surveys in the time domain, and a validation of data taken with an LMT, by comparison with data from a classical telescope, will validate their reliability. We start from a published data analysis that classified only 206 of the 18,000 astronomical objects observed with the NODO liquid mirror telescope as peculiar. A total of 29 of these 206 objects were found in the SDSS. The reliability of the NODO data can be seen by the results of the detailed analysis that, in practice, less than 0.3% of the 18,000 spectra were incorrectly identified as peculiar objects, most probably because they are variable stars. We conclude that the liquid mirror telescope gave reliable observations, comparable to those that would have been obtained with a telescope using a glass mirror.

# 1. INTRODUCTION

Liquid mirror telescopes (LMTs) are a novel type of telescope that has advantages and disadvantages with respect to telescopes that use glass mirrors. Their outstanding advantage comes from their very low cost. Inexpensive liquid mirrors with excellent surface quality have been made with mercury. Their excellent optical quality has been proven by laboratory tests (Girard & Borra 1997, Tremblay & Borra 2000). Very inexpensive zenith-pointing telescope mirrors of liquid mercury have been made with diameters as large as 6 meters (Hickson et al. 2007). (Hickson et al. 2007). discuss in details the construction, maintenance and operation of the 6 meters diameter Large Zenith Telescope (LZT) liquid mirror telescope. In particular, they discuss the mercury safety procedures and show that mercury vapors are not a problem, provided some simple safety measures are taken. LMTs have also been used for atmospheric studies (Sica & Argall 2007). The LZT has mostly been used for atmospheric studies related to adaptive optics because of the low percentage of clear nights of its Canadian location. Pfrommer & Hickson (2014) presents results of 3 years of observations with the LZT. This is the most extensive data set on sodium variability and is being used by all extremely large telescopes projects for adaptive optics development.

The main advantage of a liquid mirror comes from its low cost. A liquid mirror costs less than 1% of the cost of a glass mirror. Consequently, for a fixed zenith telescope, the cost of the entire telescope, including frame dome, corrector and detector is less than 5% of a similar telescope with a glass mirror. The simplicity of the telescope also gives advantages of low maintenance and operation costs. (Hickson et al. 2007) write that the 6-m diameter LZT can be operated routinely, easily, safely and at low cost. The LZT can be operated by a single person and only requires part-time maintenance by one person. The annual maintenance cost of the LZT is approximately $ 10,000, about 2% of the construction cost of the entire facility. The greatest disadvantage of a mercury liquid mirror is that the mirror cannot be tilted and therefore can only observe within a narrow strip of sky running through the zenith. The strip has a width limit set by the corrector. On the other hand, this inconvenient can be minimized by using active optical correctors that can correct over very large fields. Borra (1993) shows that such a corrector would allow to correct thin strips of sky to zenith distances as large as 45 degrees. Moretto & Borra (1997) describe a practical example of such a corrector. Hickson (2002) describes a corrector capable of correcting over a 1-arcmin diameter field of view up to a zenith angle of 4 degrees, thereby allowing access up to 7 per cent of the sky. Also, viscous liquid mirror can be tilted (Gagné, Borra & Ritcey 2006) and a next generation of viscous liquid mirror telescopes should allow to observe regions of the sky essentially as large as conventional survey telescopes observe,

LMTs are particularly interesting in the present era of large surveys and, in particular, surveys in the time domain. The time domain is currently a subject of great interest and the planned Large Synoptic Survey Telescope (Ivezic et al. 2014) will investigate it. It will observe the entire available sky every few nights with full operation expected in 2022. The advantage of a LMT over the LSST is that several inexpensive LMTs could be located at different sites and could observe the different strips of sky running though their zeniths night after night full time, hence allow a better time sampling. Furthermore some LMTs could use specialized instrumentation to detect very fast time variations. For example Content et al. (1989) used a 1.2-m LMT, built on the Laval university campus, to search for optical flares and flashes. They used a very inexpensive homemade telescope and its simplicity can be seen in a photograph of the telescope in Borra, Beauchemin & Lalande (1985). Their search shows that optical flares can easily be detected with a dedicated LMT and that LMTs can be built with very limited funds. As Content et al. (1989) explain in their discussion one could easily observe flares, or periodic fluctuations, with time resolutions as small as 0.3 seconds by simply observing stellar trails with a fixed CCD detector, like they did with a photographic detector. Also, Pawlowski et a. (2001) have used the NASA Orbital Debris Observatory (NODO) LMT to study the 1999 Leonid meteor shower.

The low cost of LMTs also allows to use them to carry out specialized surveys with specialized instrumentation (e.g. interference filters) optimized for particular scientific goals**.** The main purpose of the present article is to demonstrate the reliability of LMT object identifications by comparing its object identifications to Sloan Digital Sky Survey (York et al. 2000) object identifications. Information about the Sloan Digital Sky Survey can also be found in the web site (http://data.sdss3.org/).

## 2. THE NASA OPTICAL DEBRIS OBSERVATORY LIQUID MIRROR TELESCOPE SPECTRA

A search for peculiar objects was carried out by Cabanac, Borra & Beauchemin (1998), who describe in details how the data were obtained and analyzed, as well as the photometric and astrometric calibrations performed. They show that the reliability of the data is demonstrated by the fact that the star counts and color histograms obtained with the NODO liquid mirror telescope fit very well the Bahcall-Soneira model of the galaxy. This section only gives a brief summary of how the data were obtained and analyzed by Cabanac, Borra & Beauchemin (1998).

The data were obtained with the NASA Orbital debris Observatory (NODO), which was a zenith pointing telescope that used a 3-meter diameter mercury liquid mirror. It was funded and operated by NASA and active from 1995 to 2002. The NODO used a first-generation liquid mirror telescope designed to observe fast-moving space debris. Some astronomical observations were carried out with NODO (Hickson & Mulrooney 1998). A total of 34 nights of observations were used by Cabanac, Borra & Beauchemin (1998) to obtain low-resolution spectra with 11 narrowband interference filters ranging between 450 nm and 950 nm. The widths of the filters increase gradually from 20 nm at 450 nm to 45 nm at 950 nm. The filters do not overlap. Figure 1 in Cabanac, Borra & Beauchemin (1998) shows the wavelength coverage of the filters. Each night used a single filter. The detector was a 2048X2048 CCD. Tracking was done by electronically stepping the CCD pixels at the sidereal rate in the east-west direction. The integration time per night per object, given by the time it takes an object to cross the CCD surface, was 97 seconds. Many of the nights were not of photometric quality and the best 11 nights, each one using a different filter, were selected. Comparisons of images taken with NODO and the Digital Sky Survey in Cabanac, Borra & Beauchemin (1998) show an excellent agreement. Hickson & Racine (2007) carry out a quantitative evaluation of NODO (as well as LZT) images, finding that the profiles of the stellar images are mostly limited by atmospheric turbulence.

Cabanac, Borra & Beauchemin (1998) checked the reliability of the data by fitting the Bahcall-Soneira model of the Galaxy to the NODO magnitude and color counts at various Galactic latitudes. Because of the very low spectral resolution, the spectra had to be simply classified as B, F, G, K or M. Their Figure 7 shows the low resolution templates used. The templates were obtained by convolving spectra obtained from published data with the NODO filters. For stellar spectra, the templates were obtained from stellar spectra in Gunn & Stryker (1983). The galactic spectra, templates were obtained from galactic spectra in Coleman, Wu, & Weedman (1980). Quasar spectra have a broad variety of spectra and the templates were obtained from a variety of publications. A Hierarchical Clustering Analysis (HCA) was used to find objects showing peculiar spectral energy distribution. These objects are discussed in the next section. An HCA is a multidimensional clustering technique that uses a minimum-variance criterion to segregate objects in groups. It is too complex to be summarized here. A description of the techniques can be found in Beauchemin & Borra (1993), Beauchemin et al. (1991) and, in greater details in Murtagh & Heck (1987).

## 3. COMPARISON BETWEEN NODO AND SDSS SPECTRA

Table 3 in Cabanac, Borra & Beauchemin (1998) lists a total of 206 objects identified as being peculiar out of total of 18,000 objects analyzed. We use their right ascensions and declinations to find them in the data release 10 Science Archive Server of the SDSS spectroscopic survey and verify the reliability of the NODO identifications by comparing the spectra. A total of 29 of the 206 objects were found. They are listed in table I, which gives the identification number in Cabanac, Borra & Beauchemin (1998), right ascensions, declinations, SDSS spectral type identification and SDSS *g* magnitudes. The SDSS assigns spectral types by fitting spectroscopic templates. The detailed description of the SDSS spectral classifications is explained in Lee et al. (2008), The spectral types and magnitudes in Cabanac, Borra & Beauchemin (1998) are given within parenthesis.

The NODO magnitudes are measured in a narrow band filter centered at 500 nm, consequently we should not expect the magnitudes to be exactly equal to the SDSS *g* magnitudes. The g magnitudes in Cabanac, Borra & Beauchemin (1998) are obtained from a transformation equation (Equation 1 in Cabanac, Borra & Beauchemin (1998) that uses the spectrum of the blue star Hz 21 from Oke (1990) for the zero-point calibration, and the Gunn & Stryker (1983) catalog of stellar spectra for the color calibration. A least-squares fits gives the transformation equation. The NODO spectral types were obtained by fitting stellar templates to the spectra; hence a stellar spectra type was assigned to all spectra, including those classified as peculiar. Objects are classified as peculiar in Cabanac, Borra & Beauchemin (1998) if the NODO spectra differ significantly from the spectral templates. Figure 13 in Cabanac, Borra & Beauchemin (1998) shows what they mean by absorption line (a), emission lines(e), breaks (b) , variable (v), and inverted continuum (i). In the SDSS survey, the spectral types are quantized within wide spectral ranges. For example all A stars are only classified as A0 or A0p and the F stars are only classified as F2, F5 or F9. The spectral types in Cabanac, Borra & Beauchemin (1998) could only be classified as O, A, B, F, G , K and M because of the very low spectral resolution (10 samples from 500 to 950 nm). The indices within parenthesis after the SDSS spectral type identify the type of peculiarity listed in Cabanac, Borra & Beauchemin (1998). Anomalous spectral features are identified by absorption (a or sa), emission (e or se), breaks (b or sb), where s stands for strong. The numbers attached to the index give the first 2 digits of the wavelengths of the features. The (ocrs) abbreviation signifies optical counterpart of a radio source. The variable (v), and inverted continuum (i) identifications signify that the peculiarity is probably due to the fact that they are variable objects. This will correctly identify peculiar variable objects (e.g. Quasars) but also make ordinary variable objects (e.g. eclipsing binaries) wrongly show up as peculiar objects because different filters are used on different nights.

Table I shows a reasonable agreement between the NODO and SDSS spectral types. A total of 10 stars classified as A0 in the SDSS surveys were correctly classified as A stars. A total of 3 stars classified as A0 in the SDSS surveys were classified as B stars and 3 as F stars. This is a minor error since the SDSS classifies all A stars as A0 or A0p and the NODO survey classifies all B stars as B and all F stars as F. An F star was correctly identified. The worse errors were NODO 1, an A0p stars wrongly classified as O, and NODO 179, an F9 star wrongly classified as K. All of the galaxies were classified as G, K or M, which are reasonable spectral types for galaxies. Note that these are faint and distant galaxies, so that it is normal that they were identified as stars by the software in the NODO survey. NODO 103 and NODO 121 were classified as QSOs in the SDSS, NODO 118 was classified as a white dwarf in the SDSS and NODO 147 was classified as a star forming galaxy (GalaxySF) in the SDSS, which is a peculiar object, so that the NODO identification as peculiar was correct. NODO 122, while identified as a normal galaxy in the SDSS, is the optical counterpart of a radio source and was therefore correctly identified as a peculiar object. The white dwarf NODO 118 is identified as a *v* object, implying that it has several absorption features probably caused by the fact that the object is a variable star. In the case of a white dwarf, these absorption features are caused by the strong Balmer lines, so that the NODO classification is correct. We therefore see that 5/ 29 (17 %) of the objects in Table I were correctly identified as peculiar objects. On the other hand, 24/ 29 (83%) were identified as NODO peculiar objects but were identified as normal objects in the SDSS. For most of these objects, this is probably due to the fact that they are variable stars because the NODO observations were carried out in a single filter per night and therefore the spectra were generated from observations at different nights for different wavelengths. Consequently, as Cabanac, Borra & Beauchemin (1998) state in page 317: " light variations will show up as spectral variations, so we should expect that a fraction of our peculiar objects are actually variable objects . This is almost certainly the case for stellar-looking continua that show a deep absorption" . They also state: " Objects that have strong enough light variations show up as peculiar spectra, and our catalog is probably heavily contaminated by these objects ". A total of 7 of the stars in Table I was found to have absorptions and 10 were classified as probable variable stars on the basis of several absorptions in the spectrum. We see therefore that, if we exclude objects identified as probable variable objects in Cabanac, Borra & Beauchemin (1998), only 7 of the objects (24 %) in Table I were incorrectly classified as peculiar objects. This is a very small fraction of the total number of objects analyzed by Cabanac, Borra & Beauchemin (1998) since they only identified 206/18000 (1.14 %) as peculiar objects. If we exclude objects identified as probable variable objects in Cabanac, Borra & Beauchemin (1998), we see that only 0.24 x 1.16 % = 0.28 % of the observed objects were incorrectly identified as peculiar objects. This very small fraction could easily be caused by statistical random noise fluctuations , peculiar events or the fact that they are variable stars. If we consider that most of the objects wrongly classified as peculiar were given correct spectral types by template fitting, this strengthens the hypothesis that the peculiarity was caused by time variations or peculiar events. We can therefore conclude that the liquid mirror telescope performed very well.

Comparing the *g* magnitudes to the NODO magnitudes we see that essentially all of the NODO magnitudes are larger than the SDSS magnitudes and the average difference is 0.37. This is a reasonable difference if we consider that the magnitudes were observed in different spectral bands (*g* filter for SDSS and narrow band filter centered at 500 nm for NODO), consequently we should not expect the magnitudes to be exactly equal.

## 4. CONCLUSION

A search for peculiar astronomical objects carried out by Cabanac, Borra & Beauchemin (1998), who used very low resolution spectra obtained with a 3-m diameter liquid mirror telescope, found 206 candidates out of 18,000 objects observed. A search in the data release 10 Science Archive Server of the Sloan Digital Sky Survey spectroscopic survey found SDSS spectra for only 29 of the 206 candidates. While only 5 of these objects were correctly identified as peculiar objects, the erroneous identification of the majority of the other 24 objects was probably caused by the fact that they are variable stars and that the NODO spectra were obtained by observing with narrowband interference filters through several nights and a different filter for each night. If we exclude the peculiar objects identified as probable variable stars in Cabanac, Borra & Beauchemin (1998) we find that only 7 of the 29 objects were wrongly identified. It is of course possible that even some of these objects are variable stars. Variations could also be caused by peculiar events (e.g. cosmic rays or instrumental flaws). Finally, we estimate that less than 0.3 % of the total of the spectra analyzed by Cabanac, Borra & Beauchemin (1998) were incorrectly identified as peculiar objects. Consequently, considering that more than 99.7% of the objects were correctly identified , we can therefore conclude that the NODO liquid mirror telescope observations gave reliable results, comparable to those that would have been obtained with a telescope using a glass mirror.

Thus research was supported by the Natural Sciences and Engineering Research Council of Canada

TABLE I

NODO Peculiar Objects Found in the Sloan Digital Sky Survey Release 10

| ID | RA (J2000) | DEC (J2000) | Spectral type | g magnitude |
|---|---|---|---|---|
| 1 | 250.85852 | 33.020321 | A0p (O, e65-80) | 16.05 (16.65) |
| 3 | 246.54266 | 32.862266 | A0 (B. a60 e75) | 16.67 (17.19) |
| 4 | 195.58782 | 32.804821 | A0 (B. a80) | 15.50 ( 16.01) |
| 8 | 226.54567 | 32.901795 | A0 (B. v) | 16.46 (16.83) |
| 15 | 189.82898 | 33.006997 | A0 (A, a55) | 16.86 (17.24) |
| 20 | 252.14304 | 33.066146 | A0 (A, a65) | 15.68 (16.15) |
| 21 | 206.60596 | 33.000884 | A0 (A, a65-90 v) | 17.35 (17.60) |
| 29 | 256.60817 | 32.913466 | A0 (A, e60) | 18.21 (18.79) |
| 30 | 225.3717 | 32.885667 | A0 (A, a60 e65) | 15.62 (15.88) |
| 31 | 263.6761 | 32.996444 | A0 (A, e65) | 15.32 (15.61) |
| 39 | 253.83569 | 32.917722 | QSO (A,v) | 18.57 (17.83) |
| 41 | 256.81721 | 32.97104 | A0 (A, v) | 17.52 (17.55) |
| 43 | 230.79689 | 33.037321 | A0 (A, v) | 16.29 (16.63) |
| 45 | 201.24343 | 32.95722 | A0 (A, v) | 17.38 (17.71) |
| 86 | 252.15813 | 32.889858 | F5 (F, se85) | 18.22 (18.07) |
| 103 | 243.22004 | 33.016071 | QSO (F, e85 v) | 18.60 (18.61) |
| 105 | 244.0897 | 32.872679 | A0 (F. v) | 15.24 (15.75) |
| 107 | 227.51387 | 32.939793 | A0 (F, e85 v) | 17.39 (17.63) |
| 116 | 237.70989 | 33.024364 | A0 (F, sa60) | 18.13 (18.58) |
| 118 | 226.27569 | 32.999851 | WhiteDwarf (F, v) | 16.77 (17.17) |
| 121 | 233.4432 | 32.935043 | QSO(F, i) | 18.95 (18.65) |
| 122 | 249.59384 | 32.853269 | Galaxy (G. ocrs, v) | 17.88 (17.81) |
| 140 | 208.62788 | 33.039954 | Galaxy (G, sb55 b80) | 18.74 (19.05) |
| 147 | 189.01658 | 32.94324 | GalaxySF(G, sb65-85) | 18.46 (18.78) |
| 161 | 187.50677 | 33.044703 | Galaxy (G, sb55 e90) | 18.66 (19.46) |
| 179 | 252.63233 | 33.021837 | F9 (K, a70) | 17.56 (17.69) |
| 185 | 249.81253 | 32.907146 | A0 (A, a60 v) | 16.93 (16.35) |
| 187 | 187.78899 | 32.948341 | Galaxy (K, sb60) | 18.42 (18.61) |
| 206 | 188.60917 | 32.889293 | Galaxy (M, b55-75) | 17.98 (18.52) |

NOTES : I) The identification number (ID) is from Table 3 in Cabanac, Borra & Beauchemin (1998). II) The units of right ascension and declination are degrees. III) Within parentheses of the spectral types are the NODO spectral types with the anomalous spectral features identified by absorption (a or sa), emission (e or se), breaks (b or sb), where sa, se, and sb stand for strong, variable (v), and inverted continuum (i) followed by the first 2 digits of the wavelengths. The (ocrs) abbreviation signifies optical counterpart of a radio source. IV) Within parentheses of the g magnitudes are the NODO magnitudes measured with a narrow-band filter centered at 500 nm.